\documentclass[epj,twocolumn,dvipdfmx]{webofc}
\usepackage[varg]{txfonts}   
\woctitle{Powders \& Grains 2017}
\begin{document}
\title{Statistical properties of gravity-driven granular discharge flow under the influence of an obstacle}
%
%

\author{\firstname{Keita} \lastname{Endo}\inst{1}\fnsep \and \firstname{Hiroaki} \lastname{Katsuragi}\inst{1}\fnsep\thanks{\email{katsurag@eps.nagoya-u.ac.jp}}
}

\institute{Department of Earth and Environmental Sciences, Nagoya University, Furocho, Chikusa, Nagoya 464-8601, Japan}

\abstract{%
Two-dimensional granular discharge flow driven by gravity under the influence of an obstacle is experimentally investigated. A horizontal exit of width $W$ is opened at the bottom of vertical Hele-Shaw cell filled with stainless-steel particles to start the discharge flow. In this experiment, a circular obstacle is placed in front of the exit. Thus, the distance between the exit and obstacle $L$ is also an important parameter. During the discharge, granular-flow state is acquired by a high-speed camera. The bulk discharge-flow rate is also measured by load cell sensors. The obtained high-speed-image data are analyzed to clarify the particle-level granular-flow dynamics. Using the measured data, we find that the obstacle above the exit affects the granular-flow field. Specifically, the existence of obstacle results in large horizontal granular temperature and small packing fraction. This tendency becomes significant when $L$ is smaller than approximately 6$D_g$ when $W \simeq 4 D_g$, where $D_g$ is diameter of particles.
}
\maketitle
\section{Introduction}
Flow rate of gravity-driven granular discharge flow through a narrow orifice is usually constant and independent of the amount of particles above the orifice. This well-known property of granular discharge flow enables hourglass to be a good time keeper. The relation between flow rate, particle size, and orifice size is empirically developed by Beberloo et al.~\cite{Beberloo:1961uy}. This Beberloo's law is a useful relation to estimate the granular flow rate when the size of orifice is large enough. However, when the orifice size is smaller than six times of particle diameter, clogging of the flow can readily be observed~\cite{To:2001ju,Zuriguel:2014dg}. In such unstable flow regime, size distribution of avalanches has been measured and analyzed extensively~\cite{To:2005eg}. They considered Markov process for calculating the clogging probability since the avalanche size distribution obeys exponential form. Arch formation at the exit region is a main reason for the clogging. The clogging probability can be reduced when an obstacle is placed just above the orifice~\cite{Zuriguel:2011wc,Zuriguel:2014fz}. Similar effect can also be observed in various other phenomena relating to bottleneck clogging~\cite{Yanagisawa:2009gq,Frank:2011fi,Zuriguel:2014fz}. According to Ref.~\cite{Zuriguel:2011wc}, the reduction of clogging probability due to the obstacle comes from the effective pressure decrease in the region of arch formation (just above the orifice). On the other hand, Roussel et al. reported that the packing fraction in the region of arch formation is an important parameter for discussing the clogging occurrence. To reveal the principal factor governing clogging occurrence, detail analysis of granular-flow field in the discharge flow is necessary. 

Statistics of granular discharge flow has been studied by several research groups~\cite{Choi:2004gi,Moka:2005jx,Orpe:2007bv}. According to these works, velocity distribution of particle motion is not Gaussian and diffusive motion can be observed in horizontal direction. Moreover, self-similarity of velocity distribution around the exit region has also been reported~\cite{Janda:2012bp}. These features in velocity distribution show particular nature of granular discharge flow. Although they could be related to the mechanics of clogging, its detail has not yet been clarified. 

As mentioned above, an obstacle has frequently been used in order to avoid the clogging. However, the detail of velocity distribution and its role for clogging-probability reduction has not yet been revealed well. Therefore, in this study, we study particle motions in granular discharge flow under the influence of an obstacle by means of particle tracking. As a first step to discuss the obstacle effect to the granular discharge flow, we will simply focus on the velocity distribution and packing fraction at the exit region just above the orifice. 

\section{Experiment}
\begin{figure}
\centering
\includegraphics[width=.45\textwidth,clip]{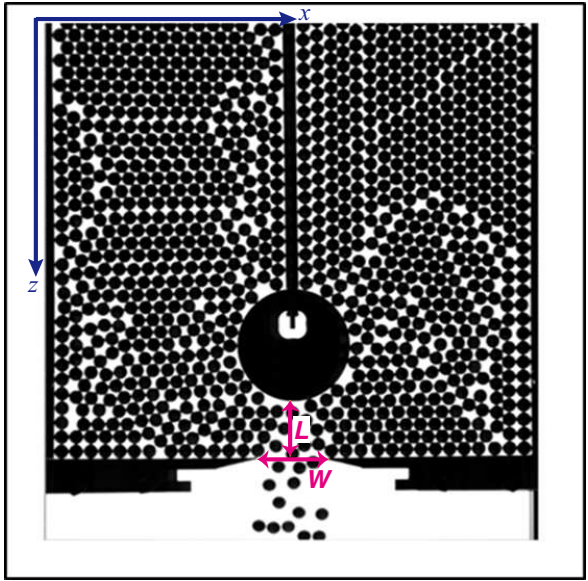}
  \caption{An example snapshot image of particles and obstacle taken by the high-speed camera. $x$ and $z$ directions correspond to horizontal and vertical directions, respectively. Width of the orifice $W$ and vertical distance between the obstacle and orifice $L$ are main controll parameters. Diameter of particles is $6.35$~mm.}
\label{fig:apparatus}
\end{figure}

A transparent (acrylic) quasi-two-dimensional cell (Hele-Shaw cell) of inner dimensions $300 \times 210 \times 6.5$~mm is vertically held and filled with stainless-steel particles of diameter $D_g=6.35$~mm. Before filling the stainless particles, a circular obstacle made with stainless steel is inserted into the cell. Diameter and thickness of the obstacle are $50$ and $6.0$~mm, respectively. The obstacle is hung on a universal testing machine (AG-X, Shimadzu) with a thin stainless-steel rod of diameter $6$~mm. After filling the particle, an exit orifice is opened at the center of cell bottom to start the discharge flow. During the discharge, flow rate is measured by load cell sensors (LMB-A, Kyowa) which quantify the mass of discharged particles. At the same time, particle motions are captured by a high-speed camera (FASTCAM SA5, Photron) with $500$~fps. The spacial resolution of this measurement is $0.26$~mm/pixel. 
An example snapshot of the particle flow is shown in Fig.~\ref{fig:apparatus}. Main control parameters in this experiment are the width of orifice $W$ and the vertical distance between obstacle and exit orifice $L$. As shown in Fig.~\ref{fig:apparatus}, vertical and horizontal directions correspond to $z$ and $x$ directions, respectively. From the high-speed video data, all particles are tracked and their velocity distributions and packing fraction are computed. 


\section{Results}
\subsection{Flow rate}
\begin{figure}
  \includegraphics[width=.48\textwidth]{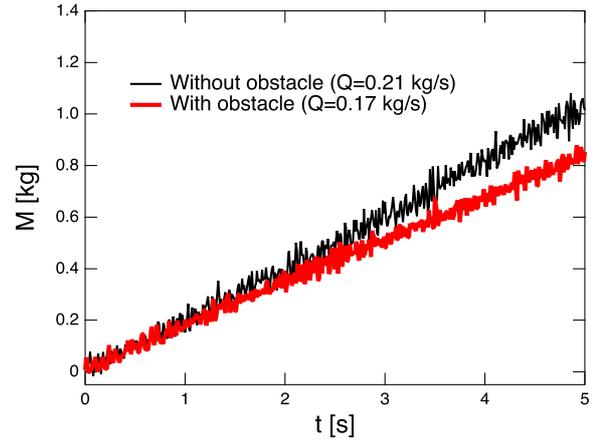}
  \caption{Discharged mass $M$ as a function of time $t$. Both cases (with- and without-obstacle) show liner relation. This means that the flow rates are constant. The value of flow rate (slope in this plot)  is reduced by introducing the obstacle.}
\label{fig:flowrate}
\end{figure}

First, the measured discharge-flow rate is shown in Fig.~\ref{fig:flowrate}. Discharged mass $M$ is shown as a function of time $t$. The important character in this plot is the linearity of these data. This linear relation indicates the constant flow rates. As can be confirmed in Fig.~\ref{fig:flowrate}, the flow rate $Q$ is reduced by introducing the obstacle. Here, in both of with- and without-obstacle cases, $W$ is fixed to $25$~mm, and $L=30$~mm is employed for the case of with-obstacle.

\begin{figure}
  \includegraphics[width=.48\textwidth]{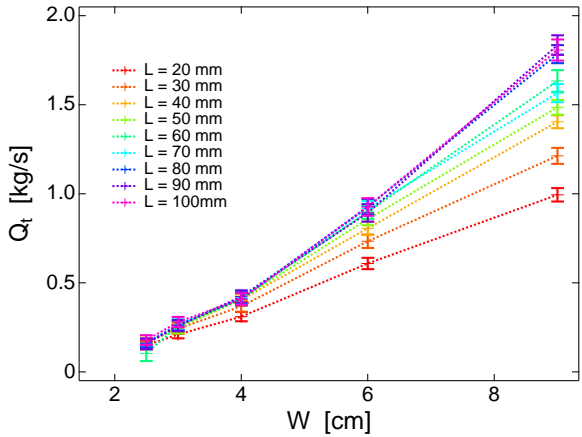}
  \caption{Orifice width $W$ dependence of flow rate $Q$ with various obstacle distance $L$. $Q$ is the increasing function of $W$ and $L$. }
\label{fig:QtWL}
\end{figure}

The value of $Q$ was systematically measured by varying $W$ and $L$. The measured results of $Q(W)$ for various $L$ are plotted in Fig.~\ref{fig:QtWL}. We can confirm that $Q$ is the increasing function of $W$ for all $L$. Clear positive correlation between $Q=|dM/dt|$ and $W$ can be seen in Fig.~\ref{fig:QtWL}. The flow rate $Q$ depends on $L$ as well. If $W$ is fixed and $L$ is varied, $Q$ becomes the increasing function of $L$. These experimental results suggest that the obstacle effectively narrows the exit region. In other words, the obstacle basically impedes discharge flow. When $W<5D_g$ is fulfilled in two-dimensional silo, the clogging could occur~\cite{To:2001ju}. Since the local arch formation is the principal reason for clogging, the narrow exit region naturally increases the risk of clogging. Although the obtained result is still preliminary, threshold value for the clogging occurrence is slightly different between horizontal ($W$) and vertical ($L$) directions. This difference might come from the anisotropy introduced by gravity. Since the current experimental system is driven by gravity, such an anisotropy might not be surprising. Furthermore, according to a recent study~\cite{Thomas:2014tm}, the clogging probability could remain finite even in very wide exit situation. Thus, the threshold might only have a practical meaning.

\subsection{Velocity distribution}
\begin{figure}
  \includegraphics[width=.48\textwidth]{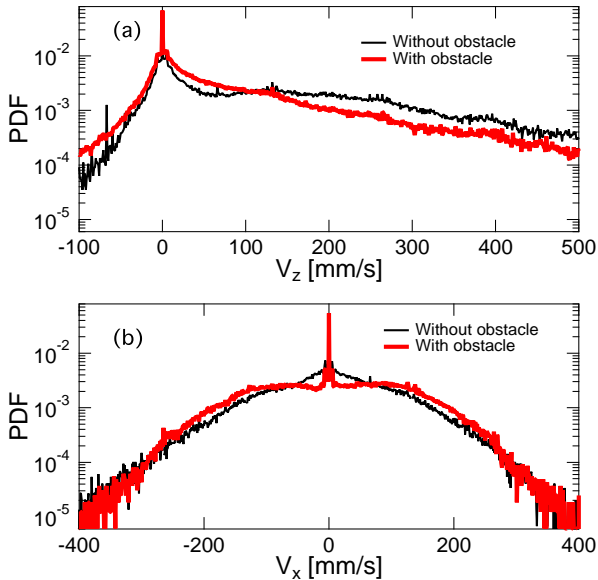}
  \caption{Probability density function (PDF) of particle velocities in (a)~vertical~[$z$] and (b)~horizontal~[$x$] directions. $W$ is fixed to $25$~mm and $L=30$~mm is used for the obstacle position. }
\label{fig:PDF}
\end{figure}

To understand the physics governing gravity-driven discharge flow, we analyzed the particle motions by using high-speed video data. Here, we restrict ourselves to the small region just above the orifice. Specifically, a small square region of width $50$~mm and height $30$~mm is particularly analyzed in the following analysis. Although the fixed square width ($50$~mm) is used in all measurements, the height is decreased to $L$ when $L$ is smaller than $30$~mm. In Fig.~\ref{fig:PDF}, probability density functions (PDF) of particle velocities are shown. The PDF of vertical component $V_z$ and horizontal component $V_x$ are presented in Fig.~\ref{fig:PDF}(a) and (b), respectively. 

Since the flow is driven by gravity, PDF of $V_z$ has an asymmetric form (Fig.~\ref{fig:PDF}(a)). The exponential-like tail (linearity in large $V_x$ regime of Fig.~\ref{fig:PDF}(a)) in PDF of $V_z$ is qualitatively consistent with the constant flow rate~(Fig.~\ref{fig:flowrate}). The difference in large $V_z$ tail part between with- and without-obstacle cases causes the difference in $Q$. One can also confirm the small amount of particles exhibiting negative $V_z$. This negative $V_z$ comes from the backscattering due to the particle collisions.

In the horizontal-velocity PDF (Fig.~\ref{fig:PDF}(b)), symmetric distribution forms can clearly be confirmed. This is natural because the gravity has nothing to do with horizontal direction. The sharp peaks around $V_x=0$ come from almost stopping particles population. Actually, the motion of particles shows stop-and-go behavior. This behavior is easily confirmed by the direct observation of high-speed video data and the similar intermittency has been observed in a recent study~\cite{Thomas:2016bt}. In addition, the qualitative form of PDF is different between with- and without-obstacle cases. PDF of $V_x$ with obstacle shows the splitting (or broadening) of the distribution. The broader part in PDF of $V_x$ stems from the enhanced particle motions. This peak splitting (or broadening) due to the existence of obstacle can be understood by two possible effects. One is the detouring of particles. If there is no obstacle on the way to exit, particles can directly approach the exit. However, the obstacle prevents this direct (nearly vertical) motion of particles. The particles must detour the obstacle to approach the exit. This detouring can directly raise the variance (i.e., granular temperature) of horizontal velocity $V_x$ distribution. The other possibility is the increase of particle collisions. As a result of detouring of particles, their collisions beneath the obstacle could be enhanced. Then, these collisions activate the particle motions. This collisional effect is also able to raise the variance (granular temperature) of $V_x$ distribution.  This is a secondary (indirect) effect of particle detouring. Note that, for $V_z$, gravity effect is much stronger than this effect.

\section{Discussion}
\begin{figure}
  \includegraphics[width=.48\textwidth]{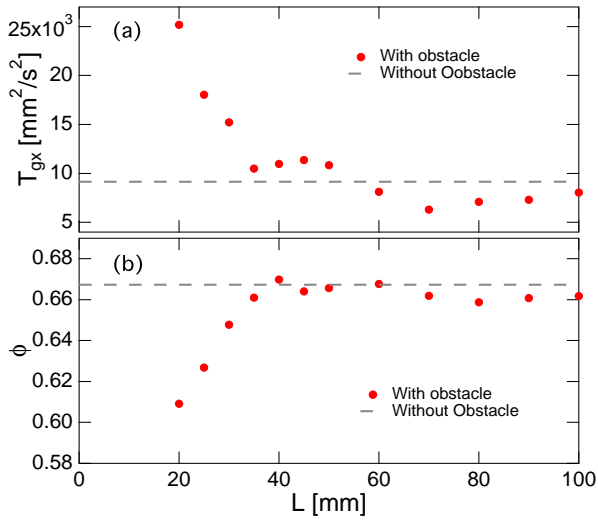}
\caption{(a)~Horizontal granular temperature $T_{gx}$ and (b)~packing fraction $\phi$ in the square region above the exit orifice (see text for the place and size of the square region). In both plots, $W$ is fixed to $25$~mm. The increase (or decrease) of $T_{gx}$ (or $\phi$) can be confirmed in small $L$($<40$~mm) regime.}
\label{fig:TgPhi}
\end{figure}

In order to characterize the granular-flow state under the influence of obstacle, we measured horizontal granular temperature $T_{gx}$ and packing fraction $\phi$ in the considered square region just above the exit orifice. Here, we simply define $T_{gx}$ by the variance of $V_x$ distribution. In Fig.~\ref{fig:TgPhi}, measured $T_{gx}$ and $\phi$ are displayed. In this measurement, $W$ was fixed to $25$~mm and $L$ was varied from $20$ to $100$~mm. Horizontal dashed lines correspond to the levels of the without-obstacle case. As can be seen in Fig.~\ref{fig:TgPhi}, $T_{gx}$ and $\phi$ become almost constant in the range of $L>40$~mm. This critical value $L\simeq 40$~mm roughly corresponds to $6D_g=38.1$~mm. And the values of $T_{gx}$ and $\phi$ are roughly identical to those in without-obstacle case in the large $L$($>40$~mm) regime. This is the reason for the small variation of $Q$ for narrow ($W=25$~mm) case observed in Fig.~\ref{fig:QtWL}. Although the data are not shown here, $Q$ with large $W$ shows clear dependence on $L$ even in relatively large $L$ regime. That is, both $W$ and $L$ are relevant to the flow characterization. However, to clearly observe the regime in which the obstacle effect is negligible, here we plot the small $W$ case in Fig.~\ref{fig:TgPhi}.

In the small $L(<40$~mm$)$ regime, $T_{gx}$ increases as $L$ is decreased~(Fig.~\ref{fig:TgPhi}(a)). This large granular temperature causes the sparse particle configuration, i.e., the effective thermal expansion might be observed. Indeed, the effective thermal expansion (small packing fraction $\phi$) in granular flow can clearly be confirmed in small $L$($<40$~mm) regime in Fig.~\ref{fig:TgPhi}(b). Perhaps, this effective thermal expansion could reduce the risk of clogging. However, the authors of Ref.~\cite{Zuriguel:2011wc} did not consider the low packing fraction is the principal reason for the clogging-probability reduction. Rather, they have considered that the pressure screening due to the backscattered particles' negative $V_z$ could be the main reason for clogging-probability reduction. The increase of population of negative $V_z$ under the influence of obstacle can surely be confirmed in this experiment, too~(Fig.~\ref{fig:PDF}(a)). Thus, we think that both effects (the low packing fraction due to enhanced particle collisions and the increase of negative $V_z$ population) could result in the reduction of risk of clogging. 

Perhaps, varying the shape of obstacle could be useful to investigate further details of clogging reduction due to the obstacle. In general, flow field around an obstacle can be affected by the shape of obstacle. Such a geometric effect might vary the clogging probability through the variation of granular-flow field. Actually, we are going to analyze the granular discharge flow with different obstacle shapes. The result will be presented elsewhere.

\section{Summary}
We performed simple granular discharge flow experiment. Flow rate $Q$, velocity distributions ($V_x$:~horizontal and $V_z$:~vertical), horizontal granular temperature $T_{gx}$, and packing fraction $\phi$ were measured and analyzed. Particularly, $T_{gx}$ and $\phi$ were measured in the vicinity of exit orifice. From the measured and analyzed results, we confirmed that the flow rate depends on both orifice width $W$ and obstacle distance $L$. Velocity distribution in vertical direction shows asymmetric form and has exponential-like tail. In addition, we found that horizontal velocity distribution shows the peak splitting (or broadening) by approaching the obstacle to the exit. To characterize this behavior, horizontal granular temperature was measured. At the same time, packing fraction was also measured. Both of them show similar increasing behaviors when the obstacle approaches as close as $L <40$~mm $\simeq 6D_g$ when $W=25$~mm $\simeq 4D_g$.

\vspace{2cm}

\section*{Acknowledgements}
We would like to acknowledge S.~Watanabe, H.~Kumagai, S.~Sirono, and T.~Morota for helpful discussion. This work has been supported by JSPS KAKENHI No.~15H03707.

\vspace{\baselineskip}

\bibliographystyle{woc}
\bibliography{Endo}

%
%
%

\end{document}